      [1999/12/01 v1.4c Il Nuovo Cimento]
\documentclass[varenna]{cimento}

\usepackage[T1]{fontenc}
\usepackage[UKenglish]{babel}
\usepackage[ansinew]{inputenc}
\usepackage{graphicx}  
\title{Accuracy of the Laser Raman system for KATRIN}

\author{M.~Schlösser \atque S.~Fischer}
\institute{Tritium Laboratory Karlsruhe, Institute for Technical Physics\\Karlsruhe Institute of Technology\\76344 Eggenstein-Leopoldshafen, Germany}

\author{M.~Hötzel \atque W.~Käfer}
\institute{Institute for Nuclear Physics\\Karlsruhe Institute of Technology\\76344 Eggenstein-Leopoldshafen, Germany}

\shortauthor{M.~Schlösser, S.~Fischer, M.~Hötzel \atque W.~Käfer}

\begin{document}

\maketitle
\begin{abstract}
The aim of the Karlsruhe Tritium Neutrino experiment (KATRIN) is the direct (model-independent) measurement of the neutrino mass. For that purpose a windowless gaseous tritium source is used, with a tritium throughput of 40 g/day. In order to reach the design sensitivity of 0.2 eV/c$^2$ (90\% C.L.) the key parameters of the tritium source, i.e. the gas inlet rate and the gas composition, have to be stabilized and monitored at the 0.1~\% level (1$\sigma$).
Any small change of the tritium gas composition will manifest itself in non-negligible effects on the KATRIN measurements; therefore, Laser Raman spectroscopy (LARA) is the method of choice for the monitoring of the gas composition because it is a non-invasive and fast in-line measurement technique.
In these proceedings, the requirements of KATRIN for statistical and systematical uncertainties of this method are discussed. An overview of the current performance of the LARA system in regard to precision will be given. In addition, two complementary approaches of intensity calibration are presented. 
\end{abstract}
\section{Introduction}
KATRIN is the next generation direct neutrino mass experiment. It aims to improve the sensitivity to $0.2\,\rm{eV\,c^{-2}}$ in the measurement of the neutrino mass (see \cite{ref:design-report} and \cite{ref:thuemmler}). The observable $m_{\bar{\nu}}^2$ is obtained by precise electron spectroscopy near the energy endpoint of the $\beta$-decay of tritium.
KATRIN will combine a Windowless Gaseous Tritium Source (WGTS) for the generation of $\beta$-electrons with a high pass filter based on the MAC-E\footnote{Magnetic Adiabatic Collimation combined with an Electrostatic Filter\cite{ref:MAC-E-Filter}} principle with a resolution of $0.93\,\rm{eV}$ for the energy analysis. 
The WGTS provides a high activity of $10^{11}$ $\beta$-decays per second. The tritium gas is injected in the center part of the source beam-tube and flows to both ends, where it is pumped off by turbomolecular pumps. The tritium injection is provided by the Inner Loop system \cite{ref:TILO-InnerLoop}. Furthermore, the reflow from the WGTS is fed back into the Inner Loop and together with the infrastructure of the hosting Tritium Laboratory Karlsruhe (TLK) \cite{ref:TLK} the tritium gas is prepared for re-injection. The injection pressure, the exit pressure at the pump ports and the WGTS beam tube temperature 
have to be kept stable on the $0.1\%$ level to achieve a stable column density in the WGTS. This is necessary to reach the KATRIN design sensitivity. However, the gas in the WGTS does not consist of pure tritium ($\rm{T_2}$), but has admixtures from the other five hydrogen isotopologues ($\rm{H_2,\,HD,\,D_2,\,HT,\,DT}$) which influence the shape and the count rate of the $\beta$-spectrum near the endpoint. Therefore, the composition of the gas has to be measured continuously. Laser Raman spectroscopy has been identified as method of choice being a non-contact, multi-species analysis technique\footnote{See \cite{ref:schloesser-Raman} for more information of Raman spectroscopy of tritium}. The Laser Raman (LARA) system developed for the KATRIN experiment has been described in detail in \cite{ref:sturm2011}. In the next section the impact of the accuracy\footnote{The terminology of ``precision'', ``trueness'' and ``accuracy'' can be found in \textsl{JCGM 200:2008 International vocabulary of metrology — Basic and general concepts and associated terms (VIM)} section 2.13-2.15~\cite{ref:JCGM}.} of the LARA system on the systematic uncertainty of KATRIN and thus on the sensitivity on the neutrino mass measurement is discussed.
\section{Impact of LARA accuracy on KATRIN}
\begin{figure}
	\centering	
	\includegraphics[width=0.60\textwidth]{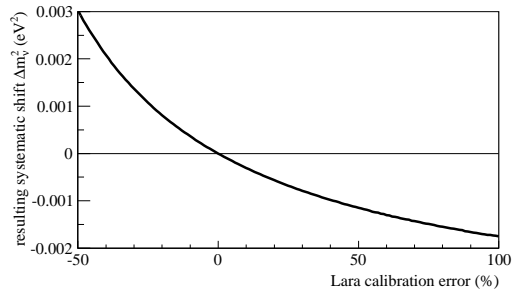}
	\caption{Dependence of the systematic shift of $m_{\bar{\nu}}^2$ on the LARA calibration error ($\sigma_{\rm{cal, LARA}}$). The LARA calibration error is defined as the percentage of misinterpretation of the calibration factor between the main components in the WGTS, $\rm{T_2}$ and $\rm{DT}$. The relation between assumed (measured) and true, relative tritium amount is $n_{\rm{assumed,T_2,rel}} = \frac{n_{\rm{true,T_2}}\cdot (1+\sigma_{\rm{cal,LARA}})}{n_{\rm{true,T_2}}\cdot (1+\sigma_{\rm{cal,LARA}})+n_{\rm{true,DT}}}$.}
		\label{fig:LARAvsmnu2_final}
\end{figure}

The gas composition of the WGTS influences the activity and thus the count rate in the $\beta$-spectrum. In this case, only relative changes are of interest. Therefore in this caseonly the precision of the LARA measurement is important, but not its trueness. The precision of the LARA system is $0.1\%$ \cite{ref:design-report}. However, the situation is different when additionally taking the ro-vibrational excitations of the daughter molecules (e.g. $\rm{^3HeT^+}$) in the $\beta$-decay into account. The so-called final state distribution describes the probability for ro-vibrational excitations of the daugther molecule after the decay. These distributions affect the shape at the endpoint of the electron spectrum \cite{ref:doss2006}. The distribution differs for the different daughter isotopologues (e.g. $\rm{^3HeT^+,\,^3HeD^+,\,^3HeH^+}$).  Here, the trueness of the LARA measurement is crucial. In contrast to the precision of LARA, no design value for the trueness is yet given. For this reason the KATRIN simulation package ``Kassiopeia''\footnote{Currently in developement. Publication is in preparation.} has been used to investigate the systematic shift on the observable $m_{\bar{\nu}}^2$ for various LARA calibration uncertainties. The aimed sensitivity on the neutrino mass of KATRIN requires the systematic error of this part to be $\Delta m_{\bar{\nu}}^2 < 0.0075\,\rm{eV^2\,c^{-4}}$ \cite{ref:design-report}. The resulting correlation from the simulation results is shown in figure \ref{fig:LARAvsmnu2_final}.
The figure shows that the LARA calibration does influence the systematic shift of the neutrino mass only moderately ($\Delta m_{\bar{\nu}}^2 \leq 0.003\,\rm{eV^2c^{-4}}$). However, the current estimation of the systematic error in the theoretical description of the final-state distribution is $0.006\,\rm{eV^2\,c^{-4}}$ \cite{ref:design-report}. The error from the theoretical description and the contribution from the LARA-calibration error have to be added in quadrature. Realistic LARA calibration errors $\sigma_{\rm{cal,LARA}}\ll50\%$ would only lead to slight increases in this combined error. 

In the next section the current status of the LARA system is presented with regard to its accuracy (precision and trueness).
\section{Status of the accuracy KATRIN Laser Raman system}
\subsection{Achieved precision}
In measurements at the test tritium loop LOOPINO a precision of $<0.1\%$  was reached within $250\,\rm{s}$ acquisition time under conditions similar to KATRIN operation~\cite{ref:fischer11}. Latest measurements show that acquisition times of $t<100\,\rm{s}$ with precisions better than $0.1\%$ are possible. Therefore, the KATRIN requirements can be easily fulfilled.
\subsection{Achieved trueness}
The trueness of a system is generally linked to the quality of its calibration. For the calibration of the KATRIN LARA system two complementary calibration approaches are currently pursued.
The first method is a classical method in which calibration samples with well-known mixtures are used. However, the accuracy of this method strongly depends on the knowledge of the sample composition. In the case of hydrogen isotopologues, the preparation of gas samples is very demanding towards an accurate knowledge of the reaction constants and mechanism. A dedicated gas mixing device has been installed at TLK which allows the preparation of gas mixtures of non-radioactive hydrogen isotopologues ($\rm{H_2,\,D_2,\,HD}$). The results from the calibration campaigns show a consistent picture and are very promising. A calibration error of $5\%$ seems to be feasible. 
The second method for the calibration of the LARA system uses theoretical \textsl{ab-initio} Raman line strength functions. Calculations from quantum mechanics define the Raman cross-sections for each hydrogen isotopologue. 
However, these values haven't been experimentally verified until now. A possibility for the verification are so-called depolarization measurements. In these measurements the ratio of isotropically and anisotropically scattered light is determined. 
An extensive campaign of the measurement of the depolarization ratio of all six hydrogen isotopologues has been finished recently at the TLK with great success. 
The obtained depolarization ratios are in agreement with the theoretical calculation within the experimental uncertainty~\cite{ref:LeRoy}. Publications are prepared on the measurement results and novel techniques.
\section{Conclusion}
The performed simulations on the influence of the LARA calibration error on the KATRIN uncertainty show that the related systematic error is currently dominated by the theoretical description of final states and not by the LARA calibration. Therefore, the systematic studies on the effects of the final-state distributions with more detailed models will continue. In the current status, LARA will meet the requirements for precision of $0.1\%$ in acquisition times of $<100\,\rm{s}$. The two complementary calibration approaches promise calibrations with trueness better than $5\%$ which will be further investigated.

\end{document}